\documentclass[preprintnumbers,unsortedaddress,superscriptaddress,notitlepage]{revtex4}
\usepackage{amsfonts}
\usepackage{amsmath}
\usepackage{hyperref}
\usepackage{amssymb}
\usepackage[english]{babel}
\usepackage{graphicx}
\usepackage{epsfig}
\usepackage{bm}
\usepackage{longtable}
\usepackage{verbatim}
\usepackage{longtable}
\usepackage{cleveref}

\crefname{equation}{Eq.}{Eqs.}
\newcommand{\mathsym}[1]{{}}

\newcommand{\emp}{\begin{equation}}
\newcommand{\fin}{\end{equation}}
\newcommand{\empn}{\begin{equation*}}
\newcommand{\finn}{\end{equation*}}

\newcommand{\bea}{\begin{eqnarray}}
\newcommand{\eea}{\end{eqnarray}}
\newcommand{\eger}{\begin{gather}}
\newcommand{\fger}{\end{gather}}
\newcommand{\egn}{\begin{gather*}}
\newcommand{\fgn}{\end{gather*}}
\newcommand{\bit}{\begin{itemize}}
\newcommand{\eit}{\end{itemize}}

\input{tcilatex}
\preprint{\bf PO-TH 14/O5}
\preprint{\bf USM-TH-323}

\begin{document}

\title{Fermion masses and mixings in an $SU(5)$ grand unified model with an
extra flavor symmetry}
\author{Miguel D. Campos}
\email{miguel.campos@postgrado.usm.cl}
\affiliation{{\small Universidad T\'ecnica Federico Santa Mar\'{\i}a and Centro Cient%
\'{\i}fico-Tecnol\'ogico de Valpara\'{\i}so}\\
Casilla 110-V, Valpara\'{\i}so, Chile}
\author{A. E. C\'arcamo Hern\'andez}
\email{antonio.carcamo@usm.cl}
\affiliation{{\small Universidad T\'ecnica Federico Santa Mar\'{\i}a and Centro Cient%
\'{\i}fico-Tecnol\'ogico de Valpara\'{\i}so}\\
Casilla 110-V, Valpara\'{\i}so, Chile}
\author{S. Kovalenko}
\email{sergey.kovalenko@usm.cl}
\affiliation{{\small Universidad T\'ecnica Federico Santa Mar\'{\i}a and Centro Cient%
\'{\i}fico-Tecnol\'ogico de Valpara\'{\i}so}\\
Casilla 110-V, Valpara\'{\i}so, Chile}
\author{Iv\'an Schmidt}
\email{ivan.schmidt@usm.cl}
\affiliation{{\small Universidad T\'ecnica Federico Santa Mar\'{\i}a and Centro Cient%
\'{\i}fico-Tecnol\'ogico de Valpara\'{\i}so}\\
Casilla 110-V, Valpara\'{\i}so, Chile}
\author{Erik Schumacher}
\email{erik.schumacher@tu-dortmund.de}
\affiliation{{\small Fakult\"at f\"ur Physik, Technische Universit\"at Dortmund}\\
D-44221 Dortmund, Germany}

\begin{abstract}
We propose a model based on the $SU(5)$ grand unification with an extra $%
A_{4}\otimes Z_{2}\otimes Z_{2}^{\prime }\otimes Z_{2}^{\prime \prime
}\otimes U\left( 1\right) _{f}$ flavor symmetry, which accounts for the
pattern of the SM fermion masses and mixings. 
The observed hierarchy of charged fermion masses and quark mixing matrix
elements arises from a generalized Froggatt-Nielsen mechanism triggered by 
a scalar $\mathbf{24}$ representation of $SU(5)$ charged under the global $%
U(1)_{f}$ and acquiring a VEV at the GUT scale. 
The light neutrino masses are generated via a radiative seesaw mechanism
with a single heavy Majorana neutrino and neutral scalars running in the
loops. The model predictions for both quark and lepton sectors are in good
agreement with the experimental data. 
The model predicts an effective Majorana neutrino mass, relevant for
neutrinoless double beta decay, with values $m_{\beta \beta }=$ 4 and 50 meV
for the normal and the inverted neutrino spectrum, respectively. The model
also features a suppression of CP violation in neutrino oscillations, a low
scale for the heavy Majorana neutrino (few TeV) and, due to the unbroken $%
Z_{2}$ symmetry, a natural dark matter candidate. 
\end{abstract}

\maketitle


\section{Introduction}

The great success of the Standard Model (SM) in the 
description of electroweak phenomena, recently confirmed with the LHC
discovery of the Higgs boson, nevertheless leaves many unresolved problems. 
Among the most pressing are the smallness of neutrino masses, 
the puzzling pattern of fermion masses and mixings, and the existence of the
three families of quarks and leptons. 
In the search for a solution of these problems various extensions of the SM
with additional flavor symmetries have been proposed in the literature (for
a review see, e.g., Refs. \cite{FlavorSymmRev,Altarelli:2002hx,Fritzsch:1999ee}). 
Historically, some of these symmetries were hinted at by %
the tribimaximal (TBM) ansatz for the leptonic mixing matrix, 
\begin{equation}
U_{\text{TBM}}=%
\begin{bmatrix}
\sqrt{\dfrac{2}{3}} & \dfrac{1}{\sqrt{3}} & 0 \\ 
-\dfrac{1}{\sqrt{6}} & \dfrac{1}{\sqrt{3}} & -\dfrac{1}{\sqrt{2}} \\ 
-\dfrac{1}{\sqrt{6}} & \dfrac{1}{\sqrt{3}} & \dfrac{1}{\sqrt{2}}%
\end{bmatrix}%
.  \label{TBM-ansatz}
\end{equation}%
leading to the neutrino mixing angles $(\sin ^{2}\theta _{12})_{\text{TBM}%
}=1/3$, $(\sin ^{2}\theta _{23})_{\text{TBM}}=1/2$ and $(\sin ^{2}\theta
_{13})_{\text{TBM}}=0$. However, recent measurements of a nonzero value of
the reactor mixing angle $\theta _{13}$ by the Daya Bay \cite{An:2012eh},
T2K \cite{Abe:2011sj}, MINOS \cite{Adamson:2011qu}, Double CHOOZ \cite{Abe:2011fz} and RENO \cite{Ahn:2012nd} have already ruled out the exact TBM
pattern, as shown in Tables \ref{NH} and \ref{IH} (based on Ref. \cite{Tortola:2012te}) for the normal (NH) and inverted (IH) hierarchies of the
neutrino mass spectrum. Nevertheless, the smallness of the reactor angle
still allows for the TBM to serve as a first-order approximation in the
construction of realistic models of lepton mixing based on flavor
symmetries. 

\begin{table}[tbh]
\begin{tabular}{|c|c|c|c|c|c|}
\hline
Parameter & $\Delta m_{21}^{2}$($10^{-5}$eV$^2$) & $\Delta m_{31}^{2}$($%
10^{-3}$eV$^2$) & $\left( \sin ^{2}\theta _{12}\right) _{\exp }$ & $\left(
\sin ^{2}\theta _{23}\right) _{\exp }$ & $\left( \sin ^{2}\theta
_{13}\right) _{\exp }$ \\ \hline
Best fit & $7.62$ & $2.55$ & $0.320$ & $0.613$ & $0.0246$ \\ \hline
$1\sigma $ range & $7.43-7.81$ & $2.46-2.61$ & $0.303-0.336$ & $0.573-0.635$
& $0.0218-0.0275$ \\ \hline
$2\sigma $ range & $7.27-8.01$ & $2.38-2.68$ & $0.29-0.35$ & $0.38-0.66$ & $%
0.019-0.030$ \\ \hline
$3\sigma $ range & $7.12-8.20$ & $2.31-2.74$ & $0.27-0.37$ & $0.36-0.68$ & 
\\ \hline
\end{tabular}%
\caption{Range for experimental values of neutrino mass squared splittings
and leptonic mixing parameters, taken from Ref. \protect\cite{Tortola:2012te}%
, for the case of normal hierarchy.}
\label{NH}
\end{table}
\begin{table}[tbh]
\begin{tabular}{|c|c|c|c|c|c|}
\hline
Parameter & $\Delta m_{21}^{2}$($10^{-5}$eV$^2$) & $\Delta m_{13}^{2}$($%
10^{-3}$eV$^2$) & $\left( \sin ^{2}\theta _{12}\right) _{\exp }$ & $\left(
\sin ^{2}\theta _{23}\right) _{\exp }$ & $\left( \sin ^{2}\theta_{13}\right)
_{\exp }$ \\ \hline
Best fit & $7.62$ & $2.43$ & $0.320$ & $0.600$ & $0.0250$ \\ \hline
$1\sigma $ range & $7.43-7.81$ & $2.37-2.50$ & $0.303-0.336$ & $0.569-0.626$
& $0.0223-0.0276$ \\ \hline
$2\sigma $ range & $7.27-8.01$ & $2.29-2.58$ & $0.29-0.35$ & $0.39-0.65 $ & $%
0.020-0.030$ \\ \hline
$3\sigma $ range & $7.12-8.20$ & $2.21-2.64$ & $0.27-0.37$ & $0.37-0.67$ & $%
0.017-0.033$ \\ \hline
\end{tabular}%
\caption{Range for experimental values of neutrino mass squared splittings
and leptonic mixing parameters, taken from Ref. \protect\cite{Tortola:2012te}%
, for the case of inverted hierarchy.}
\label{IH}
\end{table}

Since the mixing patterns of leptons and quarks are significantly different,
it is challenging to implement a unique symmetry, able to describe the small
quark mixing angles and the (two) large leptonic ones at the same time. %

Grand unified theories (GUTs), endowed with global flavor symmetries, may be
an appropriate setup for a unified description of 
the masses and mixings 
of leptons and quarks. This is motivated by the fact that leptons and quarks
are members of the same multiplets of the GUT group, which relates their
masses and mixings. 
\cite{Marzocca:2011dh,Antusch:2013kna}. 
Various GUT models with flavor symmetries have been proposed in the
literature 
\cite%
{Chen:2013wba,King:2012in,Meloni:2011fx,BhupalDev:2012nm,Babu:2009nn,Babu:2011mv,Gomez-Izquierdo:2013uaa,Antusch:2010es,Hagedorn:2010th,Ishimori:2008fi,Patel:2010hr}%
. 
For a general review see for example \cite{King:2013eh,Chen:2003zv}.

In this paper we propose a version of the $SU(5)$ GUT model with an
additional global flavor symmetry group 
\mbox{$A_4\times Z_2\times
Z_2^{\prime}\times Z_2^{\prime\prime}\times U(1)_{f}$}. 
It involves a horizontal symmetry $U_{f}(1)$, allowing us to naturally
introduce 
the fermion mass hierarchies through a generalized Froggatt-Nielsen
mechanism \cite{Wang:2011ub}. The discrete symmetry groups $A_{4}$ and three
different $Z_{2}$ are needed in order to reproduce the specific patterns of
mass matrices in the quark and lepton sectors. 
%


The embedding of the model in a nonminimal $SU$(5) GUT requires a
significant extension of the scalar sector. The particular role of each
additional scalar field and the corresponding particle assignments under the
symmetry group of the model are explained in details in Sec. \ref{model}. On
the other hand, in analogy to Ref. \cite{Hernandez:2013dta}, we consider
only one additional right-handed neutrino $N_{R}$ in order to explain the
masses and mixings in the neutrino sector. The light neutrino masses are
generated in our model through a radiative seesaw mechanism, in which
neutrinos receive their masses only from radiative corrections at one-loop
level. The smallness of the neutrino masses is a natural consequence of the
small one-loop contributions and the quadratic dependence on the neutrino
Yukawa couplings. In contrast to the regular seesaw type I scenarios, the
mass of the right-handed neutrino can therefore be kept at the TeV scale. 
For a general review of the radiative seesaw we refer readers, for example,
to Ref. \cite{Ma:1998dn}, and to Ref. \cite{Ahn:2013mva} for its discussion
in the context of flavor symmetries.

Our model describes a realistic pattern of the SM fermion masses and
mixings. The model has 14 free effective parameters, which allow us to
reproduce the experimental values of 18 observables, i.e., 9 charged fermion
masses, 2 neutrino mass squared splittings, 3 lepton mixing parameters and 4
parameters of the Wolfenstein parametrization of the CKM quark mixing
matrix. 
Let us note that the similar model of Ref. \cite{Antusch:2010es}, with an $%
SU(5)$ GUT supersymmetric setup and flavor symmetries, has 14 free effective
parameters aimed at reproducing the above mentioned 18 observables.

The paper is organized as follows. In Sec. \ref{model} we outline the
proposed model. In Sec. \ref{massmix} we present our results regarding
neutrino masses and mixing, which is followed by a numerical analysis. Our
results for the quark sector, with the corresponding numerical analysis, are
presented in Sec. \ref{Quarkmixing}. We conclude with discussions and a
summary in Sec. \ref{Summary}. Some necessary facts about the $A_{4}$ group
are collected in Appendix \ref{A}.

\section{The Model}

\label{model} As is well known, the minimal $SU\left( 5\right) $ GUT \cite{Georgi:1974sy} with fermions in $\mathbf{\bar{5}}+\mathbf{10}$ and the
scalars in $\mathbf{5}+\mathbf{24}$ representations of $SU\left( 5\right) $,
suffers from various problems. In particular, it predicts wrong relations
between the down-type quark and charged lepton masses, short proton
life-time, and the unification of gauge couplings does not agree with the
values of $\alpha _{S}$, $\sin \theta _{W}$ and $\alpha _{em}$ at the $M_{Z}$
scale. There is no place in the minimal model for a nonzero neutrino mass,
in contradiction with the neutrino oscillation experiments. %
Some of these problems can be solved by an extension of the model field
content including, in particular, a scalar $\mathbf{45}$ representation of $%
SU(5)$ \cite{Georgi:1979df,Frampton:1979,Ellis:1979,Nandi:1980sd,Frampton:1980,Langacker:1980js,Kalyniak:1982pt,Giveon:1991,Dorsner:2007fy,Dorsner:2006dj,FileviezPerez:2007nh,Perez:2008ry,Khalil:2013ixa}%
. %
However, in this next-to-minimal $SU\left( 5\right) $ GUT 
the hierarchy among the fermion masses is not understood and translates to
an unexplained hierarchy among the different Yukawa couplings. This
motivates implementing a generalized Froggat-Nielsen mechanism, where the
fermion mass hierarchy is explained by a spontaneously broken group $%
U(1)_{f} $ with a special $U(1)_{f}$ charge assignment to the fields
participating in the Yukawa terms. Our model is a multi-Higgs extension of
the next-to-minimal $SU\left( 5\right) $ GUT, and the full symmetry $\mathcal{%
G}$ is broken in two subsequent steps:

\begin{equation}  \label{Group}
\begin{aligned} \mathcal{G}=SU\left( 5\right) \otimes A_{4}\otimes &
Z_{2}\otimes Z_{2}^{\prime }\otimes Z_{2}^{\prime \prime }\otimes
U(1)_{f}\\[2mm] & \Downarrow \Lambda _{GUT}\\[2mm] SU\left( 3\right)
_{C}\otimes SU & \left( 2\right) _{L}\otimes U\left( 1\right) _{Y}\otimes
Z_{2} \\[2mm] &\Downarrow \Lambda _{EW}\\[2mm] SU\left( 3\right) _{C}\otimes
& U\left( 1\right) _{em}\otimes Z_{2} \end{aligned}
\end{equation}
The discrete non-Abelian tetrahedral symmetry group $A_4$, the group of even
permutations of four objects, 
is the smallest group with one three-dimensional and three distinct
one-dimensional irreducible representations (irreps), naturally
accommodating the three families of fermions. In the literature this group
has been extensively studied in the context of the flavor problem and
neutrino physics (cf. \cite{Ahn:2013mva,Ishimori:2012fg,Morisi:2013qna}). %
The role of the other symmetry group factors of $\mathcal{G}$ 
will be explained in what follows. 

In the present model the fermion sector is extended by introducing only one
additional field, a Majorana neutrino $N_{R}$ which is a singlet under the
SM group. 
The three families of left- and right-handed fermions, corresponding to the $%
\overline{\mathbf{5}}$ irrep of $SU\left( 5\right) $, are unified into an $%
A_{4}$ triplet in order to have one Yukawa term for the interaction with the
right-handed neutrino $N_{R}$, analogously to Ref. \cite{Hernandez:2013dta}.
The three families of left- and right-handed fermions accomodated into a $%
\mathbf{10}$ irrep of $SU\left( 5\right) $ are assigned to the three
different $A_{4}$ singlets $\mathbf{1,1^{\prime },1^{\prime \prime }}$. The
only right-handed SM singlet neutrino $N_{R}$ of our model is assigned to
the $\mathbf{1}$ of $A_{4}$ in order for its Majorana mass term be invariant
under this symmetry. The presence of this term is crucial for our
construction, as explained below. Note that neither the $\mathbf{1^{\prime }}
$ nor $\mathbf{1^{\prime \prime }}$ singlet representations of $A_{4}$
satisfy this condition, as can be seen from the multiplication rules in Eq. (%
\ref{A4-singlet-multiplication}). The fermion assignments under the group $%
\mathcal{G}=SU(5)\otimes A_{4}\otimes Z_{2}\otimes Z_{2}^{\prime }\otimes
Z_{2}^{\prime \prime }\otimes U(1)_{f}$ are 
\begin{equation}
\psi ^{i}=\left( \psi ^{i\left( 1\right) },\psi ^{i\left( 2\right) },\psi
^{i\left( 3\right) }\right) \sim \left( \overline{\mathbf{5}}\mathbf{,3,}%
1,1,-1,Q_{\overline{\mathbf{5}}}^{\left( \psi \right) }\right) ,\hspace{0.5cm%
}\hspace{0.5cm}N_{R}\sim \left( \mathbf{1,1},-1,1,-1,0\right) .
\end{equation}%
\begin{equation}
\Psi _{ij}^{\left( 1\right) }\sim \left( \mathbf{10,1,}1,1,1,Q_{\mathbf{10}%
}^{\left( 1\right) }\right) ,\hspace{0.5cm}\Psi _{ij}^{\left( 2\right) }\sim
\left( \mathbf{10,1}^{\prime }\mathbf{,}1,1,1,Q_{\mathbf{10}}^{\left(
2\right) }\right) ,\hspace{0.5cm}\Psi _{ij}^{\left( 3\right) }\sim \left( 
\mathbf{10,1}^{\prime \prime }\mathbf{,}1,1,1,Q_{\mathbf{10}}^{\left(
3\right) }\right) ,\hspace{0.5cm}i,j=1,2,3,4,5.
\end{equation}%
More explicitly, the fermions are accommodated as \cite{Binetruy:2007} 
\begin{equation}
\Psi _{ij}^{\left( f\right) }=\frac{1}{\sqrt{2}}\left( 
\begin{array}{ccccc}
0 & u_{3}^{\left( f\right) c} & -u_{2}^{\left( f\right) c} & -u_{1}^{\left(
f\right) } & -d_{1}^{\left( f\right) } \\ 
-u_{3}^{\left( f\right) c} & 0 & u_{1}^{\left( f\right) c} & -u_{2}^{\left(
f\right) } & -d_{2}^{\left( f\right) } \\ 
u_{2}^{\left( f\right) c} & -u_{1}^{\left( f\right) c} & 0 & -u_{3}^{\left(
f\right) } & -d_{3}^{\left( f\right) } \\ 
u_{1}^{\left( f\right) } & u_{2}^{\left( f\right) } & u_{3}^{\left( f\right)
} & 0 & -l^{\left( f\right) c} \\ 
d_{1}^{\left( f\right) } & d_{2}^{\left( f\right) } & d_{3}^{\left( f\right)
} & l^{\left( f\right) c} & 0%
\end{array}%
\right) _{L},\hspace{1.5cm}f=1,2,3\hspace{1.5cm}i,j=1,2,3,4,5.
\end{equation}

\begin{equation}
\psi ^{i\left( f\right) }=\left( d_{1}^{\left( f\right)c},d_{2}^{\left(
f\right)c },d_{3}^{\left( f\right)c },l^{\left( f\right)},-\nu _{f}\right)_L.
\end{equation}

Here the subscripts correspond to the different quark colors, while the
superscript $f$ refers to fermion families.

The scalar sector is composed of the following $SU\left( 5\right) $
representations: one $\mathbf{24}$, one $\mathbf{45}$, seven $\mathbf{5}$'s
and six $\mathbf{1}$'s. 
One set of three $\mathbf{5}$'s and the two sets of $SU\left( 5\right) $
singlets are unified into three $A_{4}$ triplets. The remaining scalar
fields, i.e., one $\mathbf{45}$, one $\mathbf{24}$ and the remaining set of
the four $\mathbf{5}$'s, are accommodated by two trivial and two different
nontrivial $A_{4}$ singlets. Thus the $\mathcal{G}$ assignments of the
scalar fields of our model are 
\begin{equation}
\chi =\left( \chi _{1},\chi _{2},\chi _{3}\right) \sim \left( \mathbf{1,3,}%
1,-1,1,0\right) ,\hspace{1.5cm}\xi =\left( \xi _{1},\xi _{2},\xi _{3}\right)
\sim \left( \mathbf{1,3,}1,1,1,Q_{\mathbf{1}}^{\left( \xi \right) }\right) ,
\end{equation}

\begin{equation}
S_{i}=\left( S_{i}^{\left( 1\right) },S_{i}^{\left( 2\right) },S_{i}^{\left(
3\right) }\right) \sim \left( \mathbf{5,3,-}1,1,1,Q_{\mathbf{5}}^{\left(
S\right) }\right) ,
\end{equation}

\begin{equation}
H_{i}^{\left( 1\right) }\sim \left( \mathbf{5,1,}1,1,-1,Q_{\mathbf{5}%
}^{\left( 1\right) }\right) ,\hspace{1cm}H_{i}^{\left( 2\right) }\sim \left( 
\mathbf{5,1}^{\prime }\mathbf{,}1,1,1,Q_{\mathbf{5}}^{\left( 2\right)
}\right) ,\hspace{1cm}H_{i}^{\left( 3\right) }\sim \left( \mathbf{5,\mathbf{1%
}^{\prime \prime },}1,1,1,Q_{\mathbf{5}}^{\left( 3\right) }\right) ,
\end{equation}%
\begin{equation}
H_{i}^{\left( 4\right) }\sim \left( \mathbf{5,\mathbf{1},}1,1,1,Q_{\mathbf{5}%
}^{\left( 4\right) }\right) ,\hspace{1cm}\Sigma _{j}^{i}\sim \left( \mathbf{%
24,1,}1,-1,1,-\frac{1}{2}\right) ,\hspace{1cm}\Phi _{jk}^{i}\sim \left( 
\mathbf{45,1,}1,1,-1,Q_{\mathbf{45}}^{\left( \Phi \right) }\right) .
\end{equation}%
We introduce two sets of $A_{4}$ triplets $SU\left( 5\right) $ singlets in
order to separate the interactions responsible for the light neutrino masses
from those that generate the down-type quark and charged lepton masses. The $%
A_{4}$ triplet $SU\left( 5\right) $ singlet $\chi $ is the only set of
scalars which is neutral under the $U\left( 1\right) _{f}$ symmetry, while
the remaining scalars have nontrivial $U\left( 1\right) _{f}$ charges.
Notice that the two sets of $\mathbf{5}$'s, i.e., $H_{i}^{\left( h\right) }$
($h=1,2,3,4$) and $S_{i}^{\left( f\right) }$ ($f=1,2,3$) have different $%
Z_{2}$ parities. With respect to the fermion sector, only the three families
of fermions, corresponding to the $\overline{\mathbf{5}}$ irrep of $SU\left(
5\right) $, are unified into an $A_{4}$ triplet. Besides that, the three
families of fermions embedded into the $\mathbf{10}$ irrep of $SU(5)$ are
assigned to three different $A_{4}$ singlets, i.e, $\mathbf{1,1^{\prime
},1^{\prime \prime }}$. Then, in order to build the required Yukawa
interactions for charged fermions, we need the following scalars: four $%
\mathbf{5}$'s assigned to $A_{4}$ singlets (two of them assigned to a $A_{4}$
trivial singlets and the other ones assigned to $A_{4}$ nontrivial
singlets), one $\mathbf{45}$ assumed to be a trivial $A_{4}$ singlet $%
\mathbf{\mathbf{1}}$, the $SU\left( 5\right) $ singlet $A_{4}$ triplet $\xi $
and the scalar field $\Sigma $ in the $\mathbf{24}$ representation of $%
SU\left( 5\right) $. As previously mentioned, having scalar fields in the $%
\mathbf{45}$ representation of $SU\left( 5\right) $ is crucial in order to
get the correct mass relations of down-type quarks and charged leptons.
Concerning the breakdown of the group $\mathcal{G}$ in Eq. (\ref{Group}),
the scalar field $\Sigma $ is needed to trigger the generalized
Froggatt-Nielsen mechanism responsible for generating the masses of charged
fermions via higher dimensional Yukawa terms. Besides that, the scalar field 
$\Sigma $ acquires a vacuum expectation value (VEV) at the GUT scale $%
\Lambda _{GUT}=10^{16}$ GeV and triggers the first step of symmetry breaking
in Eq. (\ref{Group}). This first step is also induced by the $A_{4}$ scalar
triplet $\xi $ acquiring a VEV at the GUT scale. The second step of symmetry
breaking, is due to the scalars $H_{i}^{\left( h\right) }$ ($h=1,2,3,4$) and 
$\Phi _{jk}^{i}$ 
acquiring VEVs at the electroweak scale. The scalar $\chi $, being an $SU(5)$
singlet, may receive its VEV at any scale below $\Lambda _{GUT}$, in
particular around TeV. The four $\mathbf{5}$'s $H_{i}^{\left( h\right) }$,
which are assigned to $A_{4}$ singlets, transform trivially under $Z_{2}$
and participate in the Yukawa interactions involving charged fermions. Since
the remaining three $\mathbf{5}$'s $S_{i}$ are unified into an $A_{4}$
triplet and transform nontrivially under $Z_{2}$, they participate in the
Yukawa interactions with the right-handed neutrino $N_{R}$. In analogy to
Ref. \cite{Hernandez:2013dta} we assume that the $Z_{2}$ symmetry is not
affected by the electroweak symmetry breaking. Therefore, the $A_{4}$
triplet $S_{i}$ does not acquire a VEV and consequently neutrinos do not
receive masses at tree-level. The preserved $Z_{2}$ discrete symmetry also
allows for stable dark matter candidates, as in Refs. \cite{Ma:2006km,Ma:2006fn}. In our model they are either the lightest neutral
component of the $SU(2)$ doublet component of $S_{i}$ or the right-handed
Majorana neutrino $N_{R}$. We do not address this question in the present
paper. As in Ref. \cite{Hernandez:2013dta}, the scalar $\chi $ generates a
neutrino mass matrix texture compatible with the experimentally observed
deviation from the TBM pattern. As we will explain in the following, the
neutrino mass matrix texture generated via the one-loop seesaw mechanism is
mainly due to the VEV of this scalar $\langle \chi \rangle =\Lambda _{int}$,
which is assumed to be much larger than the scale of the electroweak
symmetry breaking $\Lambda _{int}\gg \Lambda _{EW}=246$ GeV and at the same
time much lower than the GUT scale $\Lambda _{int}\ll \Lambda _{GUT}=10^{16}$
GeV. This, along with the assumption that the scalars (excepting $\chi $)
are charged under $U(1)_{f}$, leads to a mixing matrix that is TBM to a good
approximation. 
The $Z_{2}^{\prime }$ discrete symmetry is also an important ingredient of
our approach. 
Once it is imposed, it forbids the terms in the scalar potential involving
odd powers of $\chi $. This results in a reduction of the number of free
model parameters and selects a particular direction of symmetry breaking in
the group space. Also, as will be shown in Sec. \ref{Quarkmixing}, due to
the $A_{4}$ assignment, the top quark gets its mass mainly from $H^{(3)}$. 
The $Z_{2}^{\prime }$ symmetry is broken after the $A_{4}$ scalar triplet $%
\chi $ field acquires a nonvanishing VEV. The symmetry $Z_{2}^{\prime \prime
}$ guaranties that the scalars giving the dominant contribution to the
masses to the down-type quarks and the charged leptons are different from
those providing masses to the up-type quarks. 
This is crucial for keeping realistic lepton mixing (sf. Ref \cite{Hernandez:2013dta}). The fact that down-type quarks and charged leptons are
unified into a $\overline{\mathbf{5}}$ irrep of $SU(5)$ will result in a
trivial contribution to the quark mixing from the down-type quark sector.
Thus, the quark mixing will arise solely from the up-type quark sector as
shown in detail in Sec. \ref{Quarkmixing}.

Since the $A_{4}$ triplet $S_{i}$ is assumed to participate in the Yukawa
interactions with the right-handed neutrino $N_{R}$, we choose its $U(1)_{f}$
charge $Q_{\mathbf{5}}^{\left( S\right) }$ to be 
\begin{equation}
Q_{\mathbf{5}}^{\left( S\right) }=-Q_{\overline{\mathbf{5}}}^{\left( \psi
\right) }.
\end{equation}%
We consider the following VEV pattern of the scalars fields of the model.
The VEVs of the scalars $H_{i}^{\left( h\right) }$ ($h=1,2,3,4$), $%
S_{i}^{\left( f\right) }$ ($f=1,2,3$) and $\Sigma _{j}^{i}$ are 
\begin{equation}
\left\langle H_{i}^{\left( h\right) }\right\rangle =v_{H}^{\left( h\right)
}\delta _{i5},\hspace{1.5cm}\left\langle S_{i}^{\left( f\right)
}\right\rangle =v_{S}^{\left( f\right) }\delta _{i5},\hspace{1.5cm}f=1,2,3,%
\hspace{1.5cm}h=1,2,3,4,
\end{equation}%
\begin{equation}
\left\langle \Sigma _{j}^{i}\right\rangle =v_{\Sigma }\,diag\left( 1,1,1,-%
\frac{3}{2},-\frac{3}{2}\right) ,\hspace{1.5cm}i,j=1,2,3,4,5.
\end{equation}%
It is worth mentioning that the VEV pattern for the $\Sigma $ field, which
is consistent with the minimization conditions of the scalar potential,
follows from the general group theory of spontaneous symmetry breakdown \cite{Li:1973mq}.

The requirement that $Z_{2}$ is preserved implies, according to the field
assignment given above, that 
\begin{equation}
v_{S}^{\left( f\right) }=0,\hspace{1.5cm}f=1,2,3.
\end{equation}%
For the VEVs of the neutral components of the $A_{4}$ triplet scalars $\chi $
and $\xi $ we assume 
\begin{equation}
v_{\chi _{1}}=-v_{\chi _{3}}=\frac{v_{\chi }}{\sqrt{2}},\hspace{1.5cm}%
v_{\chi _{2}}=0,\hspace{1.5cm}v_{\xi _{1}}=v_{\xi _{2}}=v_{\xi _{3}}=\frac{%
v_{\xi }}{\sqrt{3}}.  \label{VEVdirections}
\end{equation}%
Here $v_{H}^{\left( h\right) }\sim \Lambda _{EW}=v=246$ GeV ($h=1,2,3$) and $%
v_{\chi }=\Lambda _{int}$. We also assume $v_{\xi }=\Lambda _{GUT}$. The
choice of directions in the $A_{4}$ space, given by Eq. (\ref{VEVdirections}%
), is justified by the observation that they describe a natural solution of
the scalar potential minimization equations. Indeed, in the single-field
case, $A_{4}$ invariance readily favors the $(1,1,1)$ direction over, e.g.,
the $(1,0,0)$ solution for large regions of parameter space. The vacuum $%
\left\langle \xi \right\rangle $ is a configuration that preserves a $Z_{3}$
subgroup of $A_{4}$, which has been extensively studied by many authors (see
for example Refs. \cite{Ma:2006sk,Altarelli:2005yp,Altarelli:2005yx,He:2006dk,Memenga:2013vc,Toorop:2010ex,Ahn:2012tv,Mohapatra:2012tb,Chen:2012st,Hernandez:2013dta}%
).

On the other hand, the property of the $\mathbf{45}$\ dimensional irrep of $%
SU(5)$ implies that the $\Phi _{jk}^{i}$ satisfies the following relations 
\cite{Frampton:1979,Georgi:1979df}: 
\begin{equation}
\Phi _{jk}^{i}=-\Phi _{kj}^{i},\hspace{1.5cm}\sum_{i=1}^{5}\Phi _{ij}^{i}=0,%
\hspace{1.5cm}i,j,k=1,2,\cdots ,5.
\end{equation}

Consequently, the only allowed nonzero VEVs of $\Phi _{jk}^{i}$ are 
\begin{equation}
\left\langle \Phi _{p5}^{p}\right\rangle =-\frac{1}{3}\left\langle \Phi
_{45}^{4}\right\rangle =v_{\Phi },\hspace{1.5cm}\left\langle \Phi
_{j5}^{i}\right\rangle =v_{\Phi }\left( \delta _{j}^{i}-4\delta
_{4}^{i}\delta _{j}^{4}\right) ,\hspace{1.5cm}i,j=1,2,3,4,5,\hspace{1.5cm}%
p=1,2,3,5.
\end{equation}

With the above particle content, the following renormalizable $\tciLaplace
_{Y}$ and higher-dimensional $\tciLaplace _{Y}^{\left( NR\right) }$ Yukawa
terms arise: 
\begin{equation}
\tciLaplace _{Y}=\lambda _{\nu }\left( \psi ^{i}S_{i}\right) _{\mathbf{1}%
}N_{R}+M_{N}\overline{N}_{R}N_{R}^{c}+H.c.,  \label{ly1}
\end{equation}%
\begin{eqnarray}
\tciLaplace _{Y}^{\left( NR\right) } &=&\frac{\alpha _{1}}{\Lambda }\left( 
\frac{\Sigma _{l}^{k}\Sigma _{k}^{l}}{\Lambda ^{2}}\right) ^{a_{\mathbf{1}%
}}\left( \psi ^{i}\xi \right) _{\mathbf{1}}H^{j\left( 1\right) }\Psi
_{ij}^{\left( 1\right) }+\frac{\alpha _{2}}{\Lambda }\left( \frac{\Sigma
_{l}^{k}\Sigma _{k}^{l}}{\Lambda ^{2}}\right) ^{a_{\mathbf{2}}}\left( \psi
^{i}\xi \right) _{\mathbf{\mathbf{1}^{\prime \prime }}}H^{j\left( 1\right)
}\Psi _{ij}^{\left( 2\right) }+\frac{\alpha _{3}}{\Lambda }\left( \frac{%
\Sigma _{l}^{k}\Sigma _{k}^{l}}{\Lambda ^{2}}\right) ^{a_{\mathbf{3}}}\left(
\psi ^{i}\xi \right) _{\mathbf{\mathbf{1}^{\prime }}}H^{j\left( 1\right)
}\Psi _{ij}^{\left( 3\right) }  \notag \\
&&+\frac{\beta _{1}}{\Lambda }\left( \frac{\Sigma _{l}^{k}\Sigma _{k}^{l}}{%
\Lambda ^{2}}\right) ^{b_{\mathbf{1}}}\left( \psi ^{i}\xi \right) _{\mathbf{1%
}}\Phi _{i}^{jk}\Psi _{jk}^{\left( 1\right) }+\frac{\beta _{2}}{\Lambda }%
\left( \frac{\Sigma _{l}^{k}\Sigma _{k}^{l}}{\Lambda ^{2}}\right)
^{b_{2}}\left( \psi ^{i}\xi \right) _{\mathbf{\mathbf{1}^{\prime \prime }}%
}\Phi _{i}^{jk}\Psi _{jk}^{\left( 2\right) }+\frac{\beta _{3}}{\Lambda }%
\left( \frac{\Sigma _{l}^{k}\Sigma _{k}^{l}}{\Lambda ^{2}}\right)
^{b_{3}}\left( \psi ^{i}\xi \right) _{\mathbf{\mathbf{1}^{\prime }}}\Phi
_{i}^{jk}\Psi _{jk}^{\left( 3\right) }  \notag \\
&&+\varepsilon ^{ijklp}\left\{ \gamma _{12}\left( \frac{\Sigma
_{n}^{m}\Sigma _{m}^{n}}{\Lambda ^{2}}\right) ^{x_{12}}\Psi _{ij}^{\left(
1\right) }H_{p}^{\left( 3\right) }\Psi _{kl}^{\left( 2\right) }+\gamma
_{21}\left( \frac{\Sigma _{n}^{m}\Sigma _{m}^{n}}{\Lambda ^{2}}\right)
^{x_{21}}\Psi _{ij}^{\left( 2\right) }H_{p}^{\left( 3\right) }\Psi
_{kl}^{\left( 1\right) }+\gamma _{22}\left( \frac{\Sigma _{n}^{m}\Sigma
_{m}^{n}}{\Lambda ^{2}}\right) ^{x_{22}}\Psi _{ij}^{\left( 2\right)
}H_{p}^{\left( 2\right) }\Psi _{kl}^{\left( 2\right) }\right.  \notag \\
&&+\left. \gamma _{11}\left( \frac{\Sigma _{n}^{m}\Sigma _{m}^{n}}{\Lambda
^{2}}\right) ^{x_{11}}\Psi _{ij}^{\left( 1\right) }H_{p}^{\left( 4\right)
}\Psi _{kl}^{\left( 1\right) }+\gamma _{23}\left( \frac{\Sigma
_{n}^{m}\Sigma _{m}^{n}}{\Lambda ^{2}}\right) ^{x_{23}}\Psi _{ij}^{\left(
2\right) }H_{p}^{\left( 4\right) }\Psi _{kl}^{\left( 3\right) }+\gamma
_{32}\left( \frac{\Sigma _{n}^{m}\Sigma _{m}^{n}}{\Lambda ^{2}}\right)
^{x_{32}}\Psi _{ij}^{\left( 3\right) }H_{p}^{\left( 4\right) }\Psi
_{kl}^{\left( 2\right) }\right.  \notag \\
&&+\left. \gamma _{13}\left( \frac{\Sigma _{n}^{m}\Sigma _{m}^{n}}{\Lambda
^{2}}\right) ^{x_{13}}\Psi _{ij}^{\left( 1\right) }H_{p}^{\left( 2\right)
}\Psi _{kl}^{\left( 3\right) }+\gamma _{31}\left( \frac{\Sigma
_{n}^{m}\Sigma _{m}^{n}}{\Lambda ^{2}}\right) ^{x_{31}}\Psi _{ij}^{\left(
3\right) }H_{p}^{\left( 2\right) }\Psi _{kl}^{\left( 1\right) }+\gamma
_{33}\left( \frac{\Sigma _{n}^{m}\Sigma _{m}^{n}}{\Lambda ^{2}}\right)
^{x_{33}}\Psi _{ij}^{\left( 3\right) }H_{p}^{\left( 3\right) }\Psi
_{kl}^{\left( 3\right) }\right\}  \label{ly2b}
\end{eqnarray}

The subscripts $\mathbf{1,1^{\prime },1^{\prime \prime }}$ denote projecting
out the corresponding $A_{4}$ singlet in the product of the two triplets.
The lightest of the physical neutral scalar states of $H^{(1)}$, $H^{(2)}$, $%
H^{(3)}$, $H^{(4)}$ and $\Phi $ should be interpreted as the SM-like 126 GeV
Higgs observed at the LHC \cite{LHC-H-discovery}. Besides that, the
low-energy effective theory will correspond to an eight Higgs doublet model
with three scalar singlets and a light scalar color octet. 
As we will show in Sec. \ref{Quarkmixing}, the dominant contribution to the
top quark mass mainly arises from $H^{(3)}$. 
The SM-like 126 GeV Higgs also receives its main contributions from the CP
even neutral state of the $SU(2)$ doublet part of $H^{(3)}$. 
The remaining scalars are heavy and outside the LHC reach. Our model is not
predictive in the scalar sector, having numerous free uncorrelated
parameters in the scalar potential that can be adjusted to get the required
pattern of scalar masses. Therefore, the loop effects of the heavy scalars
contributing to certain observables can be suppressed by the appropriate
choice of the free parameters in the scalar potential. Fortunately, all
these adjustments do not affect the charged fermion and neutrino sector,
which is completely controlled by the fermion-Higgs Yukawa couplings and by
certain combinations of $U(1)_{f}$ charges $Q_{r}^{(i)}$ appearing in the
Yukawa terms of Eq. (\ref{ly2b}). The dimensionless couplings $\alpha _{i}$, 
$\beta _{i}$ and $\gamma _{ij}$ ($i,j=1,2,3$) in Eq. (\ref{ly2b}) are $%
\mathcal{O}(1)$ parameters and the following relations for the
Froggat-Nielsen powers are fulfilled:

\begin{eqnarray}
a_{i} &=&Q_{\mathbf{10}}^{\left( i\right) }+Q_{\overline{\mathbf{5}}%
}^{\left( \psi \right) }+Q_{\overline{\mathbf{5}}}^{\left( 1\right) }+Q_{%
\mathbf{1}}^{\left( \xi \right) },\hspace{2cm}b_{i}=Q_{\mathbf{10}}^{\left(
i\right) }+Q_{\overline{\mathbf{5}}}^{\left( \psi \right) }-Q_{\mathbf{45}%
}^{\left( \Phi \right) }+Q_{\mathbf{1}}^{\left( \xi \right) },\hspace{2cm}%
i=1,2,3.  \notag \\
x_{12} &=&x_{21}=Q_{\mathbf{10}}^{\left( 1\right) }+Q_{\mathbf{10}}^{\left(
2\right) }+Q_{\mathbf{5}}^{\left( 3\right) },\hspace{1cm}\hspace{1cm}%
x_{33}=Q_{\mathbf{10}}^{\left( 3\right) }+Q_{\mathbf{10}}^{\left( 3\right)
}+Q_{\mathbf{5}}^{\left( 3\right) },  \notag \\
x_{13} &=&x_{31}=Q_{\mathbf{10}}^{\left( 1\right) }+Q_{\mathbf{10}}^{\left(
3\right) }+Q_{\mathbf{5}}^{\left( 2\right) },\hspace{1cm}\hspace{1cm}%
x_{22}=Q_{\mathbf{10}}^{\left( 2\right) }+Q_{\mathbf{10}}^{\left( 2\right)
}+Q_{\mathbf{5}}^{\left( 2\right) },  \notag \\
x_{23} &=&x_{32}=Q_{\mathbf{10}}^{\left( 2\right) }+Q_{\mathbf{10}}^{\left(
3\right) }+Q_{\mathbf{5}}^{\left( 4\right) },\hspace{1cm}\hspace{1cm}%
x_{11}=Q_{\mathbf{10}}^{\left( 1\right) }+Q_{\mathbf{10}}^{\left( 1\right)
}+Q_{\mathbf{5}}^{\left( 4\right) },  \label{FN-powers}
\end{eqnarray}%
Furthermore, in order to relate quark masses with the quark mixing
parameters, we set 
\begin{equation}
\kappa =\frac{\Sigma _{l}^{k}\Sigma _{k}^{l}}{\Lambda ^{2}}=\frac{%
15v_{\Sigma }^{2}}{2\Lambda ^{2}}=\frac{\Lambda _{GUT}}{\Lambda }=\lambda .
\label{parameterset}
\end{equation}%
where $\lambda =0.225$ is one of the parameters in the Wolfenstein
parametrization. It is worth mentioning that the terms in the first and
second lines of Eq. (\ref{ly2b}) contribute to the masses of the down-type
quarks and charged leptons, while the remaining terms give contributions to
the up-type quark masses.

Note that in order to reproduce the nontrivial quark mixing consistent with
experimental data, the up-type quark sector requires three $\mathbf{5}$'s,
i.e., $H_{i}^{\left( 2\right) }$, $H_{i}^{\left( 3\right) }$and $%
H_{i}^{\left( 4\right) }$ irreps of $SU\left( 5\right) $ assigned to
different $A_{4}$ singlets. In the down-type quark sector, on the other
hand, only one $\mathbf{5}$ irrep $H_{i}^{\left( 1\right) }$, one $\mathbf{45%
}$ irrep $\Phi _{jk}^{i}$ and three $\mathbf{\mathbf{1}}$'s, unified in the $%
A_{4}$ triplet $\xi =(\xi _{1},\xi _{2},\xi _{3})$, are needed. As will be
shown in the next sections the same set of irreps in the up-type quark
sector would lead to the trivial Cabbibo-Kobayashi-Maskawa (CKM) mixing
matrix.

%

\section{Lepton masses and mixing}

\label{massmix}

The charged lepton mass matrix follows from Eq. (\ref{ly2b}) by using the
product rules for the $A_{4}$ group given in Appendix \ref{A}, 
\begin{equation}
M_{l}=\frac{v_{\xi }}{\sqrt{2}\Lambda }V_{lL}^{\dag }\left( 
\begin{array}{ccc}
\alpha _{1}\kappa ^{a_{1}}v_{H}^{(1)}-6\beta _{1}\kappa ^{b_{1}}v_{\Phi } & 0
& 0 \\ 
0 & \alpha _{2}\kappa ^{a_{2}}v_{H}^{(1)}-6\beta _{2}\kappa ^{b_{2}}v_{\Phi }
& 0 \\ 
0 & 0 & \alpha _{3}\kappa ^{a_{3}}v_{H}^{(1)}-6\beta _{3}\kappa
^{b_{3}}v_{\Phi }%
\end{array}%
\right) =V_{lL}^{\dag }diag\left( m_{e},m_{\mu },m_{\tau }\right) ,
\end{equation}%
with 
\begin{equation}
V_{lL}=\frac{1}{\sqrt{3}}\left( 
\begin{array}{ccc}
1 & 1 & 1 \\ 
1 & \omega & \omega ^{2} \\ 
1 & \omega ^{2} & \omega%
\end{array}%
\right) ,\hspace{2cm}\omega =e^{\frac{2\pi i}{3}}.  \label{Ml}
\end{equation}%
Since we assume that the dimensionless couplings $\alpha _{i}$ and $\beta
_{i}$ ($i=1,2,3$) are roughly of the same order of magnitude and we consider
the VEVs $v_{H}^{(1)}$ and $v_{\Phi }$ of the order of the electroweak scale 
$v\simeq 246$ GeV, the hierarchy among the charged lepton masses are
explained by different combinations of $U(1)_{f}$ charges appearing in the
Yukawa terms of Eq. (\ref{ly2b}).

\begin{figure}[tbh]
\includegraphics[width=20cm,height=22cm]{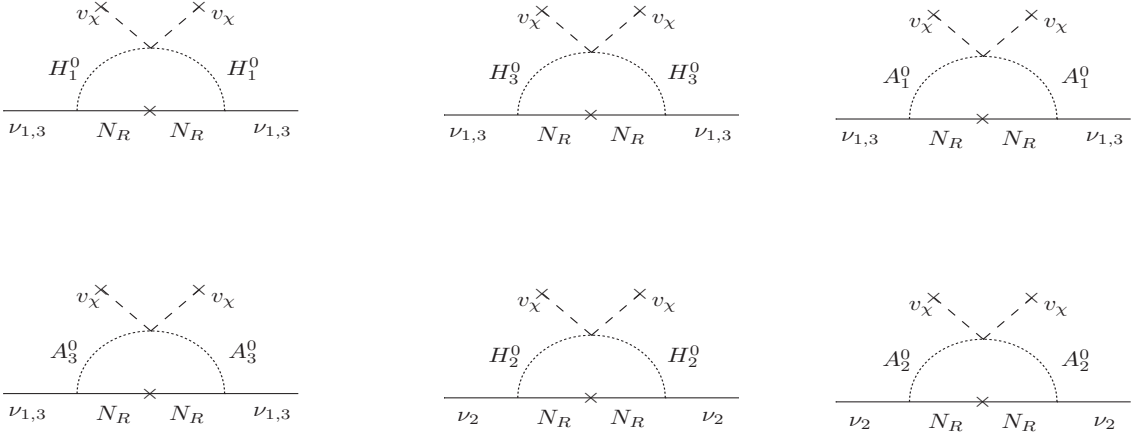} 
\vspace{-14cm}
\caption{One-loop Feynman diagrams contributing to the entries of the
neutrino mass matrix.}
\label{figMu}
\end{figure}

Since the neutral components of the scalar fields $S_{i}$ have vanishing
VEVs, the neutrino mass term does not appear at tree level, as in Ref. \cite{Hernandez:2013dta}. It arises at one-loop level in the form of a Majorana
mass term,%
\begin{equation}
-\frac{1}{2}\bar{\nu}M_{\nu }\nu ^{C}+\mbox{H.c.,}  \label{Nu-Mass-Term}
\end{equation}%
from radiative corrections involving the neutral components $H_{i}^{0}$ and $%
A_{i}^{0}$ of the $SU(2)$ doublet part of $S_{i}$ as well as the heavy
Majorana neutrino $N_{R}$ running in the internal lines of the loops. The
corresponding diagrams are shown in Fig. \ref{figMu}.\newline
\vspace{-1.5cm} 
\newline
Due to the assumption $v_{\chi }>>v$, the quartic scalar interactions
relevant for the computation of the neutrino mass matrix are given by the
terms 
\begin{eqnarray}
V\left( S,\chi \right) &=&\lambda _{1}^{\left( S\chi \right) }\left(
S_{l}S^{l}\right) _{\mathbf{1}}\left( \chi \chi \right) _{\mathbf{1}%
}+\lambda _{2}^{\left( S\chi \right) }\left[ \left( S_{l}S^{l}\right) _{%
\mathbf{1}^{\prime }}\left( \chi \chi \right) _{\mathbf{1}^{\prime \prime
}}+\left( S_{l}S^{l}\right) _{\mathbf{1}^{\prime \prime }}\left( \chi \chi
\right) _{\mathbf{1}^{\prime }}\right] +\lambda _{3}^{\left( S\chi \right)
}\left( S_{l}S^{l}\right) _{\mathbf{3s}}\left( \chi \chi \right) _{\mathbf{3s%
}}  \notag \\
&&+\lambda _{4}^{\left( S\chi \right) }\left[ e^{i\frac{\pi }{2}}\left(
S_{l}S^{l}\right) _{\mathbf{3a}}\left( \chi \chi \right) _{\mathbf{3s}}+H.c.%
\right] .
\end{eqnarray}

Following Ref \cite{Hernandez:2013dta} we choose the quartic scalar
couplings in the previous expression to be nearly universal, i.e., 
\begin{equation}
\lambda =\lambda _{2}^{\left( S\chi \right) }=\lambda _{3}^{\left( S\chi
\right) }=\lambda _{1}^{\left( S\chi \right) }-\varepsilon .
\label{univ-eps}
\end{equation}%
In practice, the coefficients need not be equal and indeed a nonzero $%
\varepsilon $ is required to generate two neutrino mass squared differences.
Hence, $\varepsilon $ parametrizes the nonuniversality of the relevant
couplings. In the approximation described above we obtain the one-loop
neutrino mass matrix in the form \cite{Hernandez:2013dta} 
\begin{equation}
M_{\nu }\simeq \left( 
\begin{array}{ccc}
Ae^{2i\psi } & 0 & A \\ 
0 & B & 0 \\ 
A & 0 & Ae^{-2i\psi }%
\end{array}%
\right) ,  \label{Mnu}
\end{equation}%
where:

\begin{equation}  \label{A-param}
\begin{aligned} A \simeq \frac{y_{\nu }^{2}}{16\pi ^{2}M_{N}}&\left\{ \left(
M_{A_{1}^{0}}^{2}-M_{A_{2}^{0}}^{2}+\frac{\varepsilon v_{\chi
}^{2}}{2}\right) \left[ D_{0}\left( \frac{M_{H_{1}^{0}}}{M_{N}}\right)
-D_{0}\left(\frac{M_{A_{1}^{0}}}{M_{N}}\right) \right] \right. \\ & +\left.
\left( M_{A_{3}^{0}}^{2}-M_{A_{2}^{0}}^{2}+\frac{ \varepsilon v_{\chi
}^{2}}{2}\right) \left[ D_{0}\left( \frac{M_{A_{3}^{0}}}{M_{N}}\right)
-D_{0}\left(\frac{M_{H_{3}^{0}}}{M_{N}}\right) \right] \right\}, \end{aligned}
\end{equation}
\begin{equation}  \label{B-param}
B \simeq \frac{\varepsilon y_{\nu }^{2}v_{\chi }^{2}}{16\pi ^{2}M_{N}}\left[
D_{0}\left( \frac{M_{H_{2}^{0}}}{M_{N}}\right) -D_{0}\left( \frac{%
M_{A_{2}^{0}}}{M_{N}}\right) \right],
\end{equation}
\begin{equation}
\tan 2\psi \simeq \frac{1}{\sqrt{\frac{9}{4}\left( \frac{%
M_{A_{3}^{0}}^{2}-M_{A_{1}^{0}}^{2}}{%
M_{A_{3}^{0}}^{2}+M_{A_{1}^{0}}^{2}-2M_{A_{2}^{0}}^{2}}\right) ^{2}-1}}\,.
\label{tan2psi}
\end{equation}%
Here $M_{H_{i}^{0}}$ and $M_{A_{i}^{0}}$ ($i=1,2,3$) are the masses of the
CP even and CP odd neutral scalars contained in the $SU(2)$ doublet
component of the $S_{i}$. %
We introduced the function \cite{Hernandez:2013mcf} 
\begin{equation}
D_{0}(x)=\frac{-1+x^{2}-\ln x^{2}}{\left( 1-x^{2}\right) ^{2}}.
\end{equation}

As can be seen from \cref{Mnu} the neutrino mass matrix depends on three
effective parameters $A$, $B$ and $\psi $, which are different combinations
of model parameters. To obtain nonvanishing neutrino masses certain
requirements need to be fulfilled. To avoid more than one massless neutrino,
the universality in the quartic couplings of the scalar potential must be
removed, which implies $\varepsilon \neq 0$ or $B\neq 0$. Additionally, to
avoid massless neutrinos at one-loop level the masses of the CP-even $%
H_{i}^{0}$ and CP-odd $A_{i}^{0}$ neutral scalars must be different. This
condition implies $A\neq 0$ and $B\neq 0$ as can be seen in %
\cref{A-param,B-param}. The parameters $A$ and $B$ constrain the neutrino
mass squared splittings, and the parameter $\psi $ constrains the
Pontecorvo-Maki-Nakagawa-Sakata (PMNS) mixing matrix, as will be shown below.


A complex symmetric Majorana mass matrix $M_{\nu }$, as in Eq. (\ref{Nu-Mass-Term}), can be diagonalized by a unitary rotation of the neutrino
fields so that 
\begin{equation*}
\nu ^{\prime }=V_{\nu }\cdot \nu \hspace{3mm}\longrightarrow \hspace{3mm}%
V_{\nu }^{\dagger }M_{\nu }(V_{\nu }^{\dagger })^{T}=diag\left(m_{\nu
_{1}},m_{\nu _{2}},m_{\nu _{3}}\right)\ \ \ \ \mbox{with}\ \ \ \ V_{\nu
}V_{\nu }^{\dagger }=\mathbf{1},
\end{equation*}%
where $m_{1,2,3}$ are real and positive. The rotation matrix has the form 
\begin{equation}
V_{\nu }=\left( 
\begin{array}{ccc}
\cos \theta & 0 & \sin \theta e^{-i\phi } \\ 
0 & 1 & 0 \\ 
-\sin \theta e^{i\phi } & 0 & \cos \theta%
\end{array}%
\right) P_{\nu },\hspace{0.5cm}\mbox{with}\hspace{0.5cm}P_{\nu }=diag\left(
e^{i\alpha _{1}/2},e^{i\alpha _{2}/2},e^{i\alpha _{3}/2}\right) ,\hspace{%
0.5cm}\theta =\pm \frac{\pi }{4},\hspace{1cm}\phi =-2\psi .  \label{tanphi}
\end{equation}%
We identify the Majorana neutrino masses and Majorana phases $\alpha_i$ for
the two possible solutions with $\theta =\pi /4,-\pi /4$ with NH and IH,
respectively. They are 
\begin{eqnarray}
\mbox{NH} &:&\theta =+\frac{\pi }{4}:\hspace{10mm}m_{\nu _{1}}=0,\hspace{10mm%
}m_{\nu _{2}}=B,\hspace{10mm}m_{\nu _{3}}=2A,\hspace{10mm}\alpha _{1}=\alpha
_{2}=0,\hspace{10mm}\alpha _{3}=\phi ,  \label{mass-spectrum-Inverted} \\%
[0.12in]
\mbox{IH} &:&\theta =-\frac{\pi }{4}:\hspace{10mm}m_{\nu _{1}}=2A,\hspace{8mm%
}m_{\nu _{2}}=B,\hspace{10mm}m_{\nu _{3}}=0,\hspace{12.5mm}\alpha
_{2}=\alpha _{3}=0,\hspace{10mm}\alpha _{1}=-\phi .
\label{mass-spectrum-Normal}
\end{eqnarray}

Note that the nonvanishing Majorana phases are $\phi $ and $-\phi $ for NH
and IH, respectively.

With the rotation matrices in the charged lepton sector $V_{lL}$, given in
Eq. (\ref{Ml}), and in the neutrino sector $V_{\nu}$, given in Eq. (\ref%
{tanphi}), we find the PMNS mixing matrix: 
\begin{equation}
U=V_{lL}^{\dag }V_{\nu }\simeq \left( 
\begin{array}{ccc}
\frac{\cos \theta }{\sqrt{3}}-\frac{e^{i\phi }\sin \theta }{\sqrt{3}} & 
\frac{1}{\sqrt{3}} & \frac{\cos \theta }{\sqrt{3}}+\frac{e^{-i\phi }\sin
\theta }{\sqrt{3}} \\ 
&  &  \\ 
\frac{\cos \theta }{\sqrt{3}}-\frac{e^{i\phi +\frac{2i\pi }{3}}\sin \theta }{%
\sqrt{3}} & \frac{e^{-\frac{2i\pi }{3}}}{\sqrt{3}} & \frac{e^{\frac{2i\pi }{3%
}}\cos \theta }{\sqrt{3}}+\frac{e^{-i\phi }\sin \theta }{\sqrt{3}} \\ 
&  &  \\ 
\frac{\cos \theta }{\sqrt{3}}-\frac{e^{i\phi -\frac{2i\pi }{3}}\sin \theta }{%
\sqrt{3}} & \frac{e^{\frac{2i\pi }{3}}}{\sqrt{3}} & \frac{e^{-\frac{2i\pi }{3%
}}\cos \theta }{\sqrt{3}}+\frac{e^{-i\phi }\sin \theta }{\sqrt{3}}%
\end{array}%
\right)P_{\nu}.  \label{PMNS}
\end{equation}%
It follows from the standard parametrization of the leptonic mixing matrix
that the lepton mixing angles are \cite{PDG}: 
\begin{equation}
\begin{aligned} \sin ^{2}\theta _{12}=\frac{\left\vert U_{e2}\right\vert
^{2}}{1-\left\vert U_{e3}\right\vert ^{2}} &= \frac{1}{2\mp \cos\phi},
\hspace{20mm} \sin ^{2}\theta _{13}=\left\vert U_{e3}\right\vert ^{2} =
\frac{1}{3}(1\pm \cos\phi), \\[3mm] &\sin ^{2}\theta _{23}=\frac{\left\vert
U_{\mu 3}\right\vert ^{2}}{1-\left\vert U_{e3}\right\vert ^{2}} = \frac{2
\mp (\cos\phi + \sqrt{3} \sin\phi)}{4 \mp 2\cos\phi}, \label{theta-ij}
\end{aligned}
\end{equation}
where the upper sign corresponds to NH ($\theta = +\pi/4$) and the lower one
to IH ($\theta = -\pi/4$). 
The PMNS matrix (\ref{PMNS}) of our model reproduces the magnitudes of the
corresponding matrix elements of the TBM ansatz (\ref{TBM-ansatz}) in the
limit $\phi =0$ (IH) and $\phi=\pi$ (NH) respectively. In both cases the
special value for $\phi$ implies that the physical neutral scalars $H_i^0$
and $A_i^0$ are degenerate in mass. Notice that the lepton mixing angles are
solely controlled by the Majorana phases $\pm\phi$, where the plus and minus
signs again correspond to NH and IH, respectively.

The Jarlskog invariant $J$ and the CP violating phase $\delta$ are given by 
\cite{PDG}: 
\begin{equation}
J=\func{Im}\left( U_{e1}U_{\mu 2}U_{e2}^{\ast }U_{\mu 1}^{\ast
}\right)\simeq-\frac{1}{6\sqrt{3}}\cos 2\theta ,\hspace{2cm}\sin \delta =%
\frac{8J}{\cos \theta _{13}\sin 2\theta _{12}\sin 2\theta _{23}\sin 2\theta
_{13}}.
\end{equation}%
Since $\theta =\pm \frac{\pi }{4}$, we predict $J\simeq 0$ and $\delta\simeq
0$ for $v_{\chi}\gg v$, implying that in our model CP violation is
suppressed in neutrino oscillations.

In the following we adjust the free parameters of our model to reproduce the
experimental values given in the \mbox{Tables \ref{NH}, \ref{IH}} and
discuss some implications of this choice of the parameters.

As seen from Eqs. (\ref{mass-spectrum-Inverted}), (\ref{mass-spectrum-Normal}%
) and (\ref{PMNS}), (\ref{theta-ij}) we have only \textit{three} effective
free parameters to fit: $\phi $, $A$ and $B$. It is noteworthy that in our
model a single parameter ($\phi $) determines all three neutrino mixing
parameters $\sin ^{2}\theta _{13}$, $\sin ^{2}\theta _{12}$ and $\sin
^{2}\theta _{23}$ as well as the Majorana phases $\alpha _{i}$. The
parameters $A$ and $B$ control the two mass squared splittings $\Delta
m_{ij}^{2}$. Therefore we actually fit only $\phi $ to adjust the values of $%
\sin ^{2}\theta _{ij}$, while $A$ and $B$ for the NH and the IH hierarchies
are simply 
\begin{eqnarray}
&&\mbox{NH}:\ m_{\nu _{1}}=0,\ \ \ m_{\nu _{2}}=B=\sqrt{\Delta m_{21}^{2}}%
\approx 9\mbox{meV},\ \ \ m_{\nu _{3}}=2A=\sqrt{\Delta m_{31}^{2}}\approx 51%
\mbox{meV};  \label{AB-Delta-NH} \\[3mm]
&&\mbox{IH}\hspace{2mm}:\ m_{\nu _{2}}=B=\sqrt{\Delta m_{21}^{2}+\Delta
m_{13}^{2}}\approx 50\mbox{meV},\ \ \ \ \ m_{\nu _{1}}=2A=\sqrt{\Delta
m_{13}^{2}}\approx 49\mbox{meV},\ \ \ m_{\nu _{3}}=0,  \label{AB-Delta-IH}
\end{eqnarray}%
as follows from Eqs. (\ref{mass-spectrum-Inverted}), (\ref%
{mass-spectrum-Normal}) and the definition $\Delta
m_{ij}^{2}=m_{i}^{2}-m_{j}^{2}$. In Eqs. (\ref{AB-Delta-NH}), (\ref%
{AB-Delta-IH}) we assumed the best-fit values of $\Delta m_{ij}^{2}$ from
Tables \ref{NH}, \ref{IH}.

Varying the model parameter $\phi $ in Eq. (\ref{theta-ij}) we have fitted
the $\sin ^{2}\theta _{ij}$ to the experimental values in Tables \ref{NH}, %
\ref{IH}. The best-fit result is 
\begin{eqnarray}
&&\mbox{NH}\ :\ \phi =-0.877\,\pi ,\ \ \ \sin ^{2}\theta _{12}\approx 0.34,\
\ \ \sin ^{2}\theta _{23}\approx 0.61,\ \ \ \sin ^{2}\theta _{13}\approx
0.0246;  \label{parameter-fit-NH} \\[3mm]
&&\mbox{IH}\hspace{2.5mm}:\ \phi =\ \ 0.12\,\pi ,\ \ \ \ \ \sin ^{2}\theta
_{12}\approx 0.34,\ \ \ \sin ^{2}\theta _{23}\approx 0.6,\ \ \ \ \,\sin
^{2}\theta _{13}\approx 0.025.  \label{parameter-fit-IH}
\end{eqnarray}

Comparing Eqs. (\ref{parameter-fit-NH}), (\ref{parameter-fit-IH}) with
Tables \ref{NH}, \ref{IH} we see that $\sin^{2}\theta _{13}$ and $\sin
^{2}\theta _{23}$ are in excellent agreement with the experimental data, for
both NH and IH, with $\sin^{2}\theta _{12}$ within a $2\sigma $ deviation
from its best fit values. It has been shown in Ref. \cite{Hernandez:2013dta}
that the solution in Eqs. (\ref{AB-Delta-NH})-(\ref{parameter-fit-IH}) does
imply neither fine-tuning nor very large values of dimensionful parameters.

With the values of the model parameters given in Eqs. (\ref{AB-Delta-NH})-(%
\ref{parameter-fit-IH}), derived from the oscillation experiments, we can
predict the amplitude for neutrinoless double beta ($0\nu\beta\beta$) decay,
which is proportional to the effective Majorana neutrino mass 
\begin{equation}
m_{\beta\beta}=\sum_jU^2_{ek}m_{\nu_k},  \label{mee}
\end{equation}
where $U^2_{ej}$ and $m_{\nu_k}$ are the PMNS mixing matrix elements and the
Majorana neutrino masses, respectively.

Then, from Eqs. (\ref{tanphi})-(\ref{PMNS}) and (\ref{AB-Delta-NH})-(\ref%
{parameter-fit-IH}), we predict the following effective neutrino masses for
both hierarchies: 
\begin{equation}
m_{\beta \beta }=\frac{1}{3}\left( B+4A\cos ^{2}\frac{\phi }{2}\right)
=\left\{ 
\begin{array}{l}
4\ \mbox{meV}\ \ \ \ \ \ \ \mbox{for \ \ \ \ NH} \\ 
50\ \mbox{meV}\ \ \ \ \ \ \ \mbox{for \ \ \ \ IH} \\ 
\end{array}%
\right.  \label{eff-mass-pred}
\end{equation}%
This is beyond the reach of the present and forthcoming $0\nu \beta \beta $
decay experiments. The presently best upper limit on this parameter $%
m_{\beta \beta }\leq 160$ meV comes from the recently quoted EXO-200
experiment \cite{Auger:2012ar} $T_{1/2}^{0\nu \beta \beta }(^{136}\mathrm{Xe}%
)\geq 1.6\times 10^{25}$ yr at 90\% C.L. This limit will be improved within
a not too distant future. The GERDA experiment \cite{Abt:2004yk,Ackermann:2012xja} is currently moving to \textquotedblleft
phase-II\textquotedblright , at the end of which it is expected to reach %
\mbox{$T^{0\nu\beta\beta}_{1/2}(^{76}{\rm Ge}) \geq 2\times 10^{26}$ yr},
corresponding to $m_{\beta \beta }\leq 100$ meV. A bolometric CUORE
experiment, using ${}^{130}Te$ \cite{Alessandria:2011rc}, is currently under
construction. Its estimated sensitivity is around $T_{1/2}^{0\nu \beta \beta
}(^{130}\mathrm{Te})\sim 10^{26}$ yr corresponding to \mbox{$m_{\beta\beta}%
\leq 50$ meV.} There are also proposals for ton-scale next-to-next
generation $0\nu \beta \beta $ experiments with $^{136}$Xe \cite%
{KamLANDZen:2012aa,Auty:2013:zz} and $^{76}$Ge \cite%
{Abt:2004yk,Guiseppe:2011me} claiming sensitivities over $T_{1/2}^{0\nu
\beta \beta }\sim 10^{27}$ yr, corresponding to $m_{\beta \beta }\sim 12-30$
meV. For recent experimental reviews, see for example Ref. \cite%
{Barabash:1209.4241} and references therein. Thus, according to Eq. (\ref%
{eff-mass-pred}) our model predicts $T_{1/2}^{0\nu \beta \beta }$ at the
level of sensitivities of the next generation or next-to-next generation $%
0\nu \beta \beta $ experiments.

\section{Quark masses and mixing}

\label{Quarkmixing}

Using Eq. (\ref{ly2b}) and the product rules for the $A_{4}$ group listed in
Appendix \ref{A}, we find the mass matrices for up- and down-type quarks in
the form 
\begin{equation}
M_{U}=\left( 
\begin{array}{ccc}
C & F & G \\ 
F & D & H \\ 
G & H & E%
\end{array}%
\right) ,  \label{MU0}
\end{equation}%
\begin{eqnarray}
M_{D} &=&\frac{v_{\xi }}{\sqrt{2}\Lambda }\left( 
\begin{array}{ccc}
\alpha _{1}\kappa ^{a_{1}}v_{H}^{(1)}+2\beta _{1}\kappa ^{b_{1}}v_{\Phi } & 0
& 0 \\ 
0 & \alpha _{2}\kappa ^{a_{2}}v_{H}^{(1)}+2\beta _{2}\kappa ^{b_{2}}v_{\Phi }
& 0 \\ 
0 & 0 & \alpha _{3}\kappa ^{a_{3}}v_{H}^{(1)}+2\beta _{3}\kappa
^{b_{3}}v_{\Phi }%
\end{array}%
\right) \left( V_{lL}^{\dag }\right) ^{T}  \notag \\
&=&diag\left( m_{d},m_{s},m_{b}\right) \left( V_{lL}^{\dag }\right) ^{T},
\label{MD}
\end{eqnarray}%
where: 
\begin{eqnarray}
F &=&2\left( \gamma _{12}+\gamma _{21}\right) \kappa ^{x_{12}}v_{H}^{\left(
3\right) },\hspace{2cm}D=4\gamma _{22}\kappa ^{x_{22}}v_{H}^{\left( 2\right)
},  \notag \\
G &=&2\left( \gamma _{13}+\gamma _{31}\right) \kappa ^{x_{13}}v_{H}^{\left(
2\right) },\hspace{2cm}C=4\gamma _{11}\kappa ^{x_{11}}v_{H}^{\left( 4\right)
}  \notag \\
H &=&2\left( \gamma _{23}+\gamma _{32}\right) \kappa ^{x_{23}}v_{H}^{\left(
4\right) },\hspace{2cm}E=4\gamma _{33}\kappa ^{x_{33}}v_{H}^{\left( 3\right)
}.  \label{C}
\end{eqnarray}%
In analogy to the leptonic sector we assume that the dimensionless couplings 
$\alpha _{i},\beta _{i},\gamma _{ij}(i,j=1,2,3)$ are roughly of the same
order of magnitude, with the VEVs $v_{H}^{(h)}(h=1,2,3,4)$ and $v_{\Phi }$
being at the electroweak scale $v\simeq 246\,$GeV. Then, the hierarchy among
the quark masses can be explained by different combinations of $U(1)_{f}$
charges shown in Eq. (\ref{FN-powers}).

The well-known hierarchy among the down-type quark masses is approximately described by 
\begin{equation}
m_{d}:m_{s}:m_{b}\approx \lambda ^{4}:\lambda ^{2}:1,
\end{equation}%
with $m_{b}\approx \lambda ^{3}m_{t}$.

To fulfill the above hierarchy, we set 
\begin{equation}
a_{1}=b_{1}=6,\hspace{1cm}a_{2}=b_{2}=4,\hspace{1cm}a_{3}=b_{3}=2,\hspace{1cm%
}v_{H}^{(1)}\sim v_{\Phi }\sim \frac{v}{\sqrt{2}}.
\end{equation}%
Here we have taken into account our previous assumption $v_{\xi }=\lambda
\Lambda $ where $\lambda =0.225$ [see Eqs. (\ref{VEVdirections}), (\ref%
{parameterset})].

Assuming that the hierarchy of charged fermion masses and quark mixing
matrix elements are explained by the Froggatt-Nielsen mechanism we adopt an
approximate universality of the dimensionless Yukawa couplings in Eq. (\ref{ly2b}). %
Specifically, we set
\begin{equation}
\beta _{1}=\beta _{3}=-\beta _{2},
\end{equation}%
so that the down-type quark and charged lepton masses will be determined by
four dimensionless parameters, i.e, $\alpha _{1}$, $\alpha _{2}$, $\alpha
_{3}$ and $\beta _{1}$. We fit these parameters to reproduce the
experimental values of the down-type quarks and charged leptons. The results
are shown in Table \ref{Observables0} for the following best-fit values of
the model parameters: 
\begin{equation}
\alpha _{1}=1.36,\hspace{1cm}\alpha _{2}=2.06,\hspace{1cm}\alpha _{3}=3.77,%
\hspace{1cm}\beta _{1}=0.18.  \label{parametersdandl}
\end{equation}%
\begin{table}[tbh]
\begin{center}
\begin{tabular}{c|l|l}
\hline\hline
Observable & Model value & Experimental value \\ \hline
$m_{d}(MeV)$ & \quad $2.91$ & \quad $2.9_{-0.4}^{+0.5}$ \\ \hline
$m_{s}(MeV)$ & \quad $57.1$ & \quad $57.7_{-15.7}^{+16.8}$ \\ \hline
$m_{b}(GeV)$ & \quad $2.73$ & \quad $2.82_{-0.04}^{+0.09}$ \\ \hline
$m_{e}(MeV)$ & \quad $0.487$ & \quad $0.487$ \\ \hline
$m_{\mu }(MeV)$ & \quad $102.8$ & \quad $102.8\pm 0.0003$ \\ \hline
$m_{\tau }(GeV)$ & \quad $1.75$ & \quad $1.75\pm 0.0003$ \\ \hline
\end{tabular}%
\end{center}
\caption{Model and experimental values of the down-type quark and charged
lepton masses (at the $M_{Z}$ scale).}
\label{Observables0}
\end{table}

As customary, we use the quark and charged lepton masses evaluated at the $%
M_Z$ scale \cite{Bora:2012tx}. 
As seen from Table \ref{Observables0} there is good agreement of the model
values for these masses with the experimental ones.

The CKM quark mixing matrix is defined as \cite{PDG} 
\begin{equation}
K=R_{U}^{\dag }R_{D}=\left( 
\begin{array}{ccc}
K_{ud} & K_{us} & K_{ub} \\ 
K_{cd} & K_{cs} & K_{cb} \\ 
K_{td} & K_{ts} & K_{tb}%
\end{array}%
\right) ,  \label{CKM}
\end{equation}%
where the rotation matrices $R_{D}$ and $R_{U}$ are derived from 
\begin{equation}
R_{U}^{\dag }M_{U}M_{U}^{\dag }R_{U}=diag\left(
m_{u}^{2},m_{c}^{2},m_{t}^{2}\right) ,\ \ \ \ \ R_{D}^{\dag
}M_{D}M_{D}^{\dag }R_{D}=diag\left( m_{d}^{2},m_{s}^{2},m_{b}^{2}\right) .
\end{equation}%
From Eq. (\ref{MD}) it follows that 
\begin{equation}
M_{D}M_{D}^{\dag }=diag\left( m_{d}^{2},m_{s}^{2},m_{b}^{2}\right) .
\end{equation}%
Thus $R_{D}=1_{3\times 3}$ and the CKM quark mixing matrix do not receive
contributions from the down-type quark sector, meaning that quark mixing
arises solely from the up-type quark sector. Thus, the CKM matrix satisfies
the following relation: 
\begin{equation}
M_{U}M_{U}^{\dag }=K^{\dag }diag\left( m_{u}^{2},m_{c}^{2},m_{t}^{2}\right)
K.  \label{eq:Mu2}
\end{equation}%
%
%
%
Now we proceed to scan over the parameters of the mass matrix for up-type
quarks looking for points where the up-type quark masses, the CKM
magnitudes, the Jarlskog invariant and the CP violating phase fit their
respective experimental values. In our model the quark CKM matrix is fully
determined by the rotation matrix of the up-type quark sector. After
scanning the parameter space, we get that a realistic pattern of the quark
masses and mixings implies that the mass matrix for up-type quarks should
satisfy
\begin{equation}
M_{U}=m_{t}\left( 
\begin{array}{ccc}
y\lambda ^{10} & f\lambda ^{9}e^{i\tau\sigma} & b\lambda ^{3} \\ 
f\lambda ^{9}e^{i\tau\sigma} & a\lambda ^{4}e^{i\sigma\left(1+\tau\right) }
& c\lambda ^{2}e^{i\sigma } \\ 
b\lambda ^{3} & c\lambda ^{2}e^{i\sigma } & de^{i\sigma }%
\end{array}%
\right) ,  \label{MU}
\end{equation}
where $y$, $a$, $b$, $c$, $d$ and $f$ are $\mathcal{O}(1)$ parameters.

%
%
%
%
To fulfill these relations, we set 
\begin{eqnarray}
x_{11} &=&10,\hspace{1cm}x_{12}=9,\hspace{1cm}x_{13}=3\hspace{1cm}x_{23}=2,%
\hspace{1cm}x_{22}=4,\hspace{1cm}x_{33}=0,  \notag \\
v_{H}^{(2)} &\sim &v_{H}^{(3)}\sim v_{H}^{(4)}\sim \frac{v}{\sqrt{2}},%
\hspace{1cm}\left\vert \gamma _{ij}\right\vert \sim \frac{1}{4},\hspace{1cm}%
i,j=1,2,3.
\end{eqnarray}%
Recall that $\Lambda =\lambda ^{-1}\Lambda _{GUT}$ and $\kappa =\lambda $,
[see Eqs. (\ref{VEVdirections}), (\ref{parameterset})]. 

Therefore, the mass matrix for up-type quarks satisfies the following
relation: 
\begin{eqnarray}
M_{U}M_{U}^{\dag } &=&m_{t}^{2}\left( 
\begin{array}{ccc}
b^{2}\lambda ^{6} & bc\lambda ^{5}e^{-i\sigma } & bd\lambda ^{3}e^{-i\sigma }
\\ 
bc\lambda ^{5}e^{i\sigma } & c^{2}\lambda ^{4} & cd\lambda ^{2} \\ 
bd\lambda ^{3}e^{i\sigma } & cd\lambda ^{2} & d^{2}%
\end{array}%
\right)  \notag \\
&&+m_{t}^{2}\left( 
\begin{array}{ccc}
\lambda ^{18}\left( f^{2}+y^{2}\lambda ^{2}\right) & f\lambda ^{13}\left(
e^{-i\sigma \tau }y\lambda ^{6}+ae^{-i\sigma }\right) & \lambda ^{11}\left(
by\lambda ^{2}+ce^{i\sigma (\tau -1)}f\right) \\ 
f\lambda ^{13}\left( e^{i\sigma \tau }y\lambda ^{6}+ae^{i\sigma }\right) & 
\lambda ^{8}\left( f^{2}\lambda ^{10}+a^{2}\right) & e^{i\sigma \tau
}\lambda ^{6}\left( bf\lambda ^{6}+ac\right) \\ 
by\lambda ^{13}+ce^{-i\sigma (\tau -1)}f\lambda ^{11} & e^{-i\sigma \tau
}\lambda ^{6}\left( bf\lambda ^{6}+ac\right) & \lambda ^{4}\left(
c^{2}+b^{2}\lambda ^{2}\right) \\ 
&  & 
\end{array}%
\right) ,  \label{MUsquared2a}
\end{eqnarray}%
Notice that the first term in Eq. (\ref{MUsquared2a}) gives the leading
contribution to $M_{U}M_{U}^{\dag }$, 
while the second one is crucial to generate the up quark mass.

From Eq. (\ref{eq:Mu2}) it follows that%
\begin{equation}
M_{U}M_{U}^{\dag }\simeq m_{t}^{2}\left( 
\begin{array}{ccc}
\left\vert K_{td}\right\vert ^{2} & K_{td}^{\dag }K_{ts} & K_{td}^{\dag
}K_{tb} \\ 
K_{ts}^{\dag }K_{td} & \left\vert K_{ts}\right\vert ^{2} & K_{ts}^{\dag
}K_{tb} \\ 
K_{tb}^{\dag }K_{td} & K_{tb}^{\dag }K_{ts} & \left\vert K_{tb}\right\vert
^{2}%
\end{array}%
\right) .  \label{MUsquared2}
\end{equation}%
Therefore we can write down 
\begin{equation}
M_{U}M_{U}^{\dag }\simeq m_{t}^{2}\left( 
\begin{array}{ccc}
W^{2}\lambda ^{6}\left[ \eta ^{2}+(\rho -1)^{2}\right]  & W^{2}\lambda
^{5}(-i\eta +\rho -1) & W\lambda ^{3}(i\eta -\rho +1) \\ 
W^{2}\lambda ^{5}(i\eta +\rho -1) & W^{2}\lambda ^{4} & -W\lambda ^{2} \\ 
W\lambda ^{3}(-i\eta -\rho +1) & -W\lambda ^{2} & 1%
\end{array}%
\right)   \label{MUsquared3},
\end{equation}%
using the Wolfenstein parameterization of the CKM matrix \cite{PDG,Wolfenstein:1983yz}: 
\begin{equation}
K\simeq \left( 
\begin{array}{ccc}
1-\frac{\lambda ^{2}}{2} & \lambda  & W\lambda ^{3}(\rho -i\eta ) \\ 
-\lambda  & 1-\frac{\lambda ^{2}}{2} & W\lambda ^{2} \\ 
W\lambda ^{3}(1-\rho -i\eta ) & -W\lambda ^{2} & 1%
\end{array}%
\right) ,
\end{equation}%
with the Wolfenstein parameters given by \cite{PDG}:%
\begin{equation}
\lambda =0.22535\pm 0.00065,\hspace{1.5cm}W=0.811_{-0.012}^{+0.022},\hspace{%
1.5cm}\overline{{\rho }}=0.131_{-0.013}^{+0.026},\hspace{1.5cm}\overline{{%
\eta }}=0.345_{-0.014}^{+0.013}  \label{wolf}
\end{equation}%
\begin{equation}
\overline{{\rho }}\simeq \rho \left( 1-\frac{{\lambda }^{2}}{2}\right) ,%
\hspace{1.5cm}\overline{{\eta }}\simeq \eta \left( 1-\frac{{\lambda }^{2}}{2}%
\right) .  \label{ckmb6}
\end{equation}%
Comparing Eqs. (\ref{MUsquared2a}) and (\ref{MUsquared3}) we find the
following relations: 
\begin{equation}
b\simeq W\sqrt{\eta ^{2}+(\rho -1)^{2}},\hspace{1.5cm}c\simeq -W,\hspace{%
1.5cm}d\simeq 1,\hspace{1.5cm}\sigma \simeq \arctan \left( -\frac{\eta }{%
1-\rho }\right) .  \label{rp1}
\end{equation}%
Since $d\simeq 1$, it follows from the previous relations that the quark
mixing in our model is described by five effective dimensionless parameters,
i.e., $b$, $c$, $\sigma $, $\tau $ and $\lambda $. The $\lambda $ parameter
in the Wolfenstein parametrization is fixed by the ratio between the grand
unification scale $\Lambda _{GUT}$ and the cutoff $\Lambda $ of our model. 
%

We fit the remaining $\mathcal{O}(1)$ parameters in Eq. (\ref{MU}) to
reproduce 
the up-type quark mass spectrum and quark mixing parameters. The results are
shown in Table \ref{Observables} for the following best-fit values: 
\begin{equation}
y=0.2,\hspace{1cm}f=0.39,\hspace{1cm}a=0.26,\hspace{1cm}\tau =2.9.
\label{parameteru}
\end{equation}

The CKM matrix in our model is in excellent agreement with the experimental
data. The agreement of our model with the experimental data is as good as in
the models of Refs. \cite{CarcamoHernandez:2010im,Bhattacharyya:2012pi,King:2013hj,Hernandez:2013hea,Hernandez:2014vta,Hernandez:2014a}
and better than, 
for example, those in Refs.~\cite{Fritzsch,Xing,FX,Matsuda,Zhou,Carcamo,CarcamoHernandez:2012xy}. 

The values of these observables as well as the up-type quark masses are
juxtaposed together with the experimental data in Table \ref{Observables}.
The experimental values of the quark masses, which are given at the $M_Z$
scale, have been taken from Ref. \cite{Bora:2012tx}, whereas the
experimental values of the CKM matrix elements and the Jarlskog invariant $J$
are taken from Ref. \cite{PDG}. As seen from Table \ref{Observables}, all
the analyzed physical parameters are in very good agreement with the
experimental data, except for $m_{u}$, and $m_c$, which reproduce the
corresponding experimental values only with order of magnitude accuracy.


\begin{table}[tbh]
\begin{center}
\begin{tabular}{c|l|l}
\hline\hline
Observable & Model value & Experimental value \\ \hline
$m_{u}(MeV)$ & \quad $5.4$ & \quad $1.45_{-0.45}^{+0.56}$ \\ \hline
$m_{c}(MeV)$ & \quad $284$ & \quad $635\pm 86$ \\ \hline
$m_{t}(GeV)$ & \quad $173.4$ & \quad $172.1\pm 0.6\pm 0.9$ \\ \hline
$\bigl|V_{ud}\bigr|$ & \quad $0.974$ & \quad $0.97427\pm 0.00015$ \\ \hline
$\bigl|V_{us}\bigr|$ & \quad $0.225$ & \quad $0.22534\pm 0.00065$ \\ \hline
$\bigl|V_{ub}\bigr|$ & \quad $0.00348$ & \quad $0.00351_{-0.00014}^{+0.00015}
$ \\ \hline
$\bigl|V_{cd}\bigr|$ & \quad $0.225$ & \quad $0.22520\pm 0.00065$ \\ \hline
$\bigl|V_{cs}\bigr|$ & \quad $0.973$ & \quad $0.97344\pm 0.00016$ \\ \hline
$\bigl|V_{cb}\bigr|$ & \quad $0.0422$ & \quad $0.0412_{-0.0005}^{+0.0011}$
\\ \hline
$\bigl|V_{td}\bigr|$ & \quad $0.00872$ & \quad $0.00867_{-0.00031}^{+0.00029}
$ \\ \hline
$\bigl|V_{ts}\bigr|$ & \quad $0.0415$ & \quad $0.0404_{-0.0005}^{+0.0011}$
\\ \hline
$\bigl|V_{tb}\bigr|$ & \quad $0.999$ & \quad $%
0.999146_{-0.000046}^{+0.000021}$ \\ \hline
$J$ & \quad $2.95\times 10^{-5}$ & \quad $(2.96_{-0.16}^{+0.20})\times
10^{-5}$ \\ \hline
$\delta $ & \quad $66^{\circ }$ & \quad $68^{\circ }$ \\ \hline\hline
\end{tabular}%
\end{center}
\caption{Model and experimental values of the up-type quark masses and CKM
parameters.}
\label{Observables}
\end{table}


\section{Conclusions}

We proposed a model based on the group $SU(5)\otimes A_{4}\otimes
Z_{2}\otimes Z_{2}^{\prime }\otimes Z_{2}^{\prime \prime }\otimes U\left(
1\right) _{f}$, which is an extension of the model of Ref. \cite%
{Hernandez:2013dta}. 
The model has in total 14 effective free parameters, which allowed us to
reproduce 18 observables, i.e., 9 charged fermion masses, 2 neutrino mass
squared splittings, 3 lepton mixing parameters and the 4 parameters of the
Wolfenstein parametrization of the CKM quark mixing matrix. The observed
hierarchy of the charged fermion masses arises from a generalized
Froggatt-Nielsen mechanism where the charged fermions get masses via
nonrenormalizable operators invariant under the gauge and flavor symmetries.
It is triggered by a scalar field $\Sigma $ in the $\mathbf{24}$
representation of $SU(5)$ charged under the global $U(1)_{f}$ symmetry and
acquiring a VEV at the GUT scale. Thus, the hierarchy of the charged fermion
masses and the quark mixing matrix elements arises as a consequence of the
power dependence of the charged fermion mass matrix elements on particular
combinations of the $U(1)_{f}$ charges.


The neutrino masses in our model arise from a radiative seesaw mechanism,
which explains their smallness, while keeping the mass of the right-handed
neutrino at comparatively low values, which could be about a few TeV. The
neutrino mixing is approximately tribimaximal due to the spontaneously
broken $A_{4}$ symmetry of the model. The experimentally observed deviation
from the TBM pattern is implemented by introducing the $SU(5)$ singlet $A_{4}
$ triplet $\chi $. Its VEV $\langle \chi \rangle \gg \Lambda _{EW}$ properly
shapes the neutrino mass matrix at the one-loop level. The model predicts
strong suppression of the CP violation in neutrino oscillations.


The predicted values of the effective Majorana neutrino mass $m_{\beta \beta
}$ for $0\nu \beta \beta $ decay are 4 and 50 meV for the normal and the
inverted neutrino spectrum, respectively.

An unbroken $Z_2$ discrete symmetry of our model also allows for stable dark
matter candidates, as in Refs. \cite{Ma:2006km,Ma:2006fn}. They could be
either the lightest neutral component of the $SU(5)$ $\mathbf{5}$-plet $S_i$
or the right-handed Majorana neutrino $N_{R}$. We do not address this
subject in the present paper.

\label{Summary}

\section*{Acknowledgments}

A.E.C.H. was partially supported by Fondecyt (Chile), Grants No. 11130115,
No. 1100582 and No. 1140390 and by DGIP internal Grant No. 111458. A.E.C.H
thanks Professor Stefan Antusch and Dr. Ivo de Mendeiros Varzielas for
useful criticism. E.S. was supported by DFG CONICYT Grant No. PA 803/7-1.
E.S. acknowledges the hospitality at UTFSM during part of this
collaboration. \appendix

\section{The product rules for $A_4$ \label{A}}

The following product rules for the $A_{4}$ group were used in the
construction of our model Lagrangian: 
\begin{eqnarray}
&&\hspace{18mm}\mathbf{3}\otimes \mathbf{3}=\mathbf{3}_{s}\oplus \mathbf{3}%
_{a}\oplus \mathbf{1}\oplus \mathbf{1}^{\prime }\oplus \mathbf{1}^{\prime
\prime },  \label{A4-singlet-multiplication} \\[3mm]
&&\mathbf{1}\otimes \mathbf{1}=\mathbf{1},\hspace{5mm}\mathbf{1}^{\prime
}\otimes \mathbf{1}^{\prime \prime }=\mathbf{1},\hspace{5mm}\mathbf{1}%
^{\prime }\otimes \mathbf{1}^{\prime }=\mathbf{1}^{\prime \prime },\hspace{%
5mm}\mathbf{1}^{\prime \prime }\otimes \mathbf{1}^{\prime \prime }=\mathbf{1}%
^{\prime }.
\end{eqnarray}%
Denoting $\left( x_{1},y_{1},z_{1}\right) $ and $\left(
x_{2},y_{2},z_{2}\right) $ as the basis vectors for two $A_{4}$-triplets $%
\mathbf{3}$, one finds

\begin{eqnarray}
&&\left( \mathbf{3}\otimes \mathbf{3}\right) _{\mathbf{1}%
}=x_{1}y_{1}+x_{2}y_{2}+x_{3}y_{3},  \label{triplet-vectors} \\
&&\left( \mathbf{3}\otimes \mathbf{3}\right) _{\mathbf{3}_{s}}=\left(
x_{2}y_{3}+x_{3}y_{2},x_{3}y_{1}+x_{1}y_{3},x_{1}y_{2}+x_{2}y_{1}\right) ,\
\ \ \ \left( \mathbf{3}\otimes \mathbf{3}\right) _{\mathbf{1}^{\prime
}}=x_{1}y_{1}+\omega x_{2}y_{2}+\omega ^{2}x_{3}y_{3}, \\
&&\left( \mathbf{3}\otimes \mathbf{3}\right) _{\mathbf{3}_{a}}=\left(
x_{2}y_{3}-x_{3}y_{2},x_{3}y_{1}-x_{1}y_{3},x_{1}y_{2}-x_{2}y_{1}\right) ,\
\ \ \left( \mathbf{3}\otimes \mathbf{3}\right) _{\mathbf{1}^{\prime \prime
}}=x_{1}y_{1}+\omega ^{2}x_{2}y_{2}+\omega x_{3}y_{3},
\end{eqnarray}%
where $\omega =e^{i\frac{2\pi }{3}}$. The representation $\mathbf{1}$ is
trivial, while the nontrivial $\mathbf{1}^{\prime }$ and $\mathbf{1}^{\prime
\prime }$ are complex conjugate to each other. Comprehensive reviews of
discrete symmetries in particle physics can be found in Refs. \cite%
{King:2013eh,Altarelli:2010gt,Ishimori:2010au,Discret-Group-Review}.

\end{document}